\documentclass[iop,referee]{emulateapj-rtx4}

\usepackage{graphicx}  % needed for figures
\usepackage{dcolumn}   % needed for some tables
\usepackage{bm}        % for math
\usepackage{amsfonts,amsmath,amssymb,mathrsfs}
\usepackage{color}
\usepackage{hyperref}
\hypersetup{
  colorlinks=true,        % false: boxed links; true: colored links
  linkcolor=blue,         % color of internal links
  citecolor=cyan,         % color of links to bibliography
}

\usepackage{natbib,times}
\newcommand{\cqg}{{CQGra}}
\renewcommand{\apjl}{{ApJL}}
\renewcommand{\nat}{{Natur}}
\renewcommand{\prd}{{PhRvD}}
\renewcommand{\prl}{{PhRvL}}

\shorttitle{Short gamma-ray bursts in the ``time-reversal'' scenario}
\shortauthors{R. Ciolfi \& D. M. Siegel}

\begin{document}

\title{Short gamma-ray bursts in the ``time-reversal'' scenario}

\author{Riccardo Ciolfi\altaffilmark{1,2} and Daniel M. Siegel\altaffilmark{2}}

\altaffiltext{1}{Physics Department, University of Trento, Via Sommarive 14, I-38123 Trento, Italy}

\altaffiltext{2}{Max Planck Institute for Gravitational Physics (Albert
  Einstein Institute), Am M\"uhlenberg 1, D-14476 Potsdam-Golm, Germany}

\email{riccardo.ciolfi@unitn.it; daniel.siegel@aei.mpg.de}

\begin{abstract}
Short gamma-ray bursts (SGRBs) are among the most luminous explosions in the universe and their origin still remains uncertain. 
Observational evidence favors the association with binary neutron star or neutron star--black hole (NS-BH) binary mergers. Leading models relate SGRBs to a relativistic jet launched by the BH-torus system resulting from the merger.
However, recent observations have revealed a large fraction of SGRB events accompanied by X-ray afterglows with durations $\sim\!10^2\!-\!10^5~\mathrm{s}$, suggesting continuous energy injection from a long-lived central engine, which is incompatible with the short ($\lesssim\!1~\mathrm{s}$) accretion timescale of a BH-torus system.
The formation of a supramassive NS, resisting the collapse on much longer spin-down timescales, can explain these afterglow durations, but leaves serious doubts on whether a relativistic jet can be launched at the merger. 
Here we present a novel scenario accommodating both aspects, where the SGRB is produced after the collapse of a supramassive NS.
Early differential rotation and subsequent spin-down emission generate an optically thick environment around the NS consisting of a photon-pair nebula and an outer shell of baryon-loaded ejecta.
While the jet easily drills through this environment, spin-down radiation diffuses outward on much longer timescales and accumulates a delay that allows the SGRB to be observed before (part of) the long-lasting X-ray signal.
By analyzing diffusion timescales for a wide range of physical parameters, we find delays that can generally reach $\sim\!10^5~\mathrm{s}$, compatible with observations. The success of this fundamental test makes this ``time-reversal" scenario an attractive alternative to current SGRB models.
\end{abstract}
		
\keywords{black hole physics --- gamma-ray burst: general --- magnetohydrodynamics (MHD) --- stars: magnetic field --- stars: neutron --- X-rays: general}

%%%%%%%%%%%%%%%%%%%%%%%%%%%%%%%%%%%%%%%%%		
\section{Introduction}
%%%%%%%%%%%%%%%%%%%%%%%%%%%%%%%%%%%%%%%%%

Merging binary neutron stars (BNSs) and neutron star--black hole
(NS-BH) systems represent the leading scenarios to
explain the phenomenology of short gamma-ray bursts (SGRBs; e.g.,
\citealt{Paczynski86,Eichler89,Narayan92,Barthelmy2005,Fox2005,Gehrels2005,Shibata06b,Rezzolla:2011,Paschalidis2014,Tanvir2013})
and are among the most promising sources of gravitational waves (GWs)
for the detection with interferometric detectors such as advanced LIGO
and Virgo \citep{Harry2010,Accadia2011}. It is generally assumed
that within $\lesssim\!100\,\text{ms}$ after the merger, a BH-torus system
forms that can power a transient relativistic jet through
accretion and thus produce the SGRB emission lasting $\lesssim2\,\text{s}$ (e.g., \citealt{Shibata06b,Rezzolla:2011,Paschalidis2014}).

\begin{figure*}[t]
\centering 
\includegraphics[angle=0,width=0.33\textwidth]{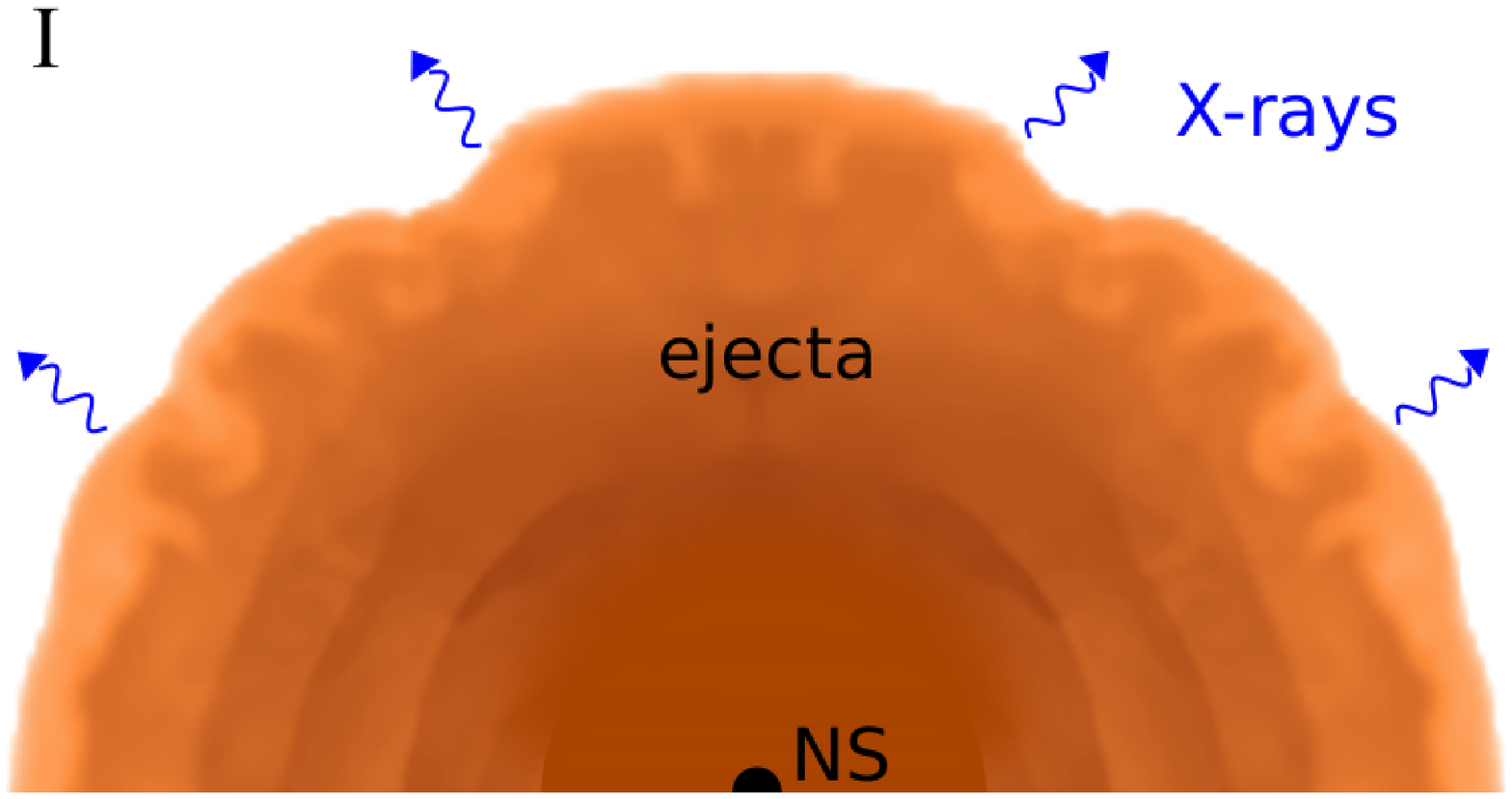}
\includegraphics[angle=0,width=0.33\textwidth]{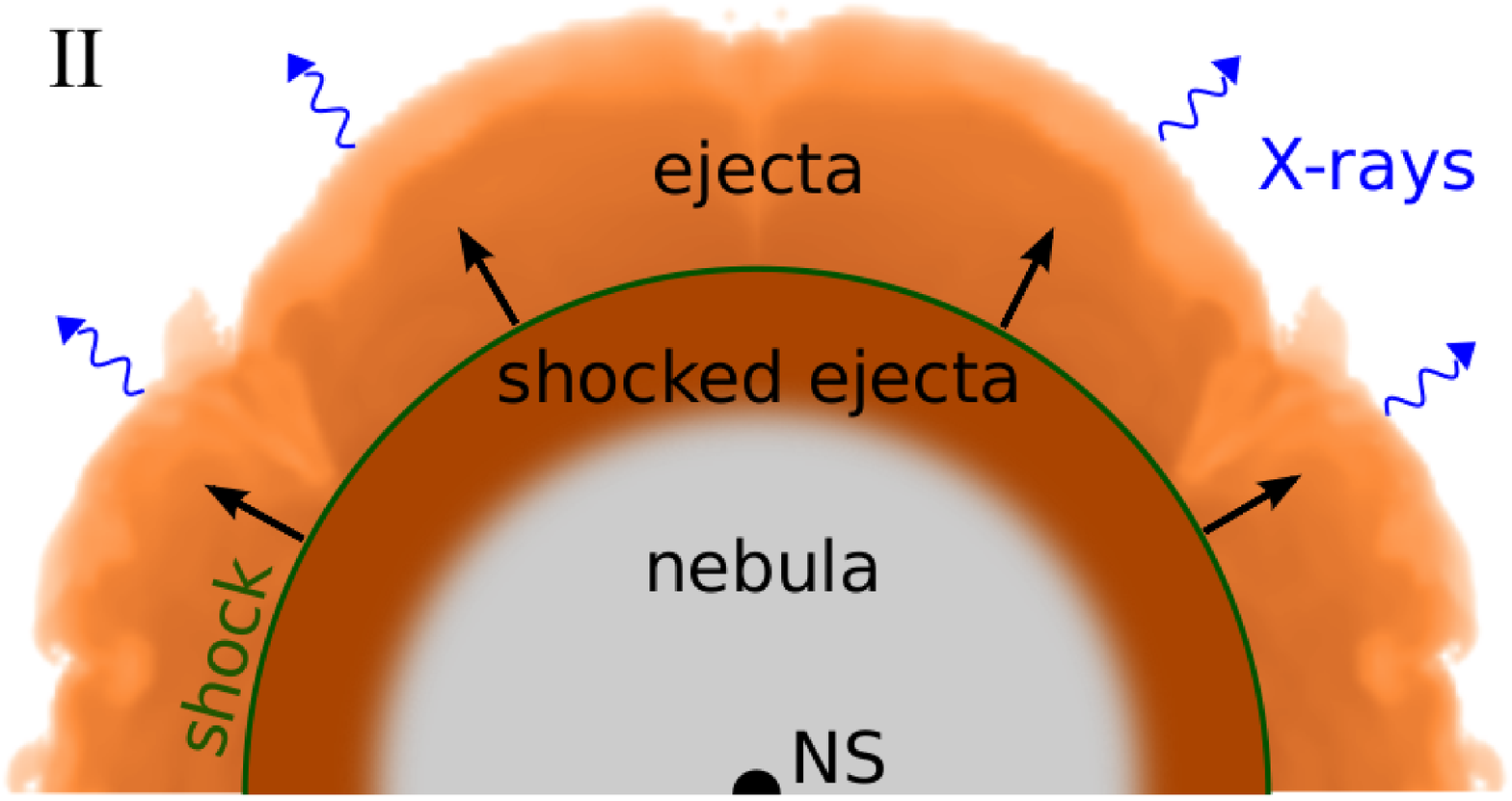}
\includegraphics[angle=0,width=0.33\textwidth]{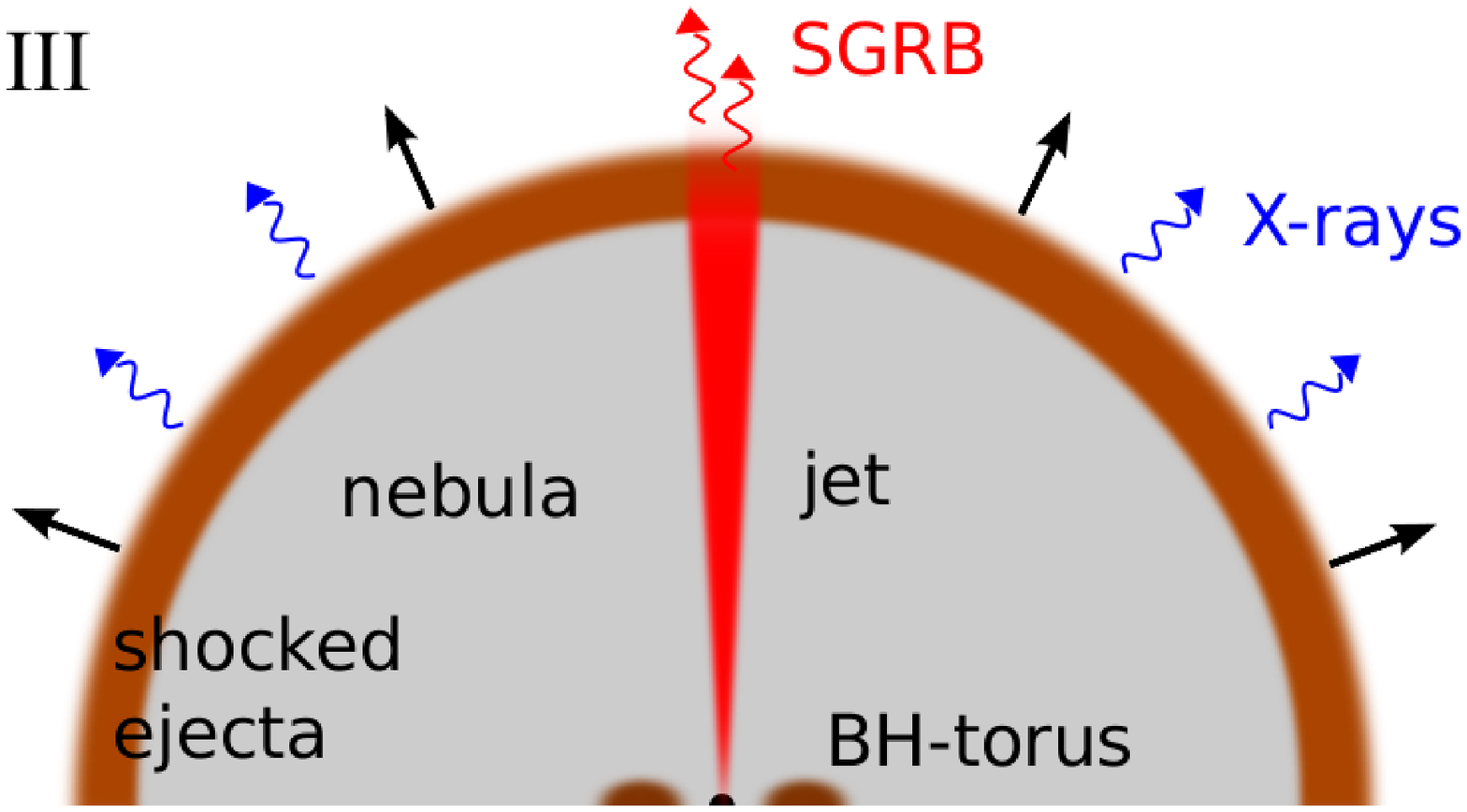}
\caption{Evolution phases: (I) The differentially rotating supramassive
  NS ejects a baryon-loaded and highly isotropic wind; (II) The
  cooled-down and uniformly rotating NS emits spin-down radiation
  inflating a photon-pair nebula that drives a shock through the
  ejecta; (III) The NS collapses to a BH, a relativistic jet drills
  through the nebula and the ejecta shell and produces the prompt
  SGRB, while spin-down emission diffuses outward on a much longer
  timescale.}
\label{fig:phenomenology} 
\end{figure*} 

Recent observations by the \emph{Swift} satellite \citep{Gehrels2004}
have revealed long-lasting, ``plateau-shaped'' X-ray
afterglows in the vast majority of observed SGRB events (e.g.,
\citealt{Rowlinson2013,Gompertz2014}). These afterglows indicate
ongoing energy injection by a central engine on timescales up to
$\sim\!10^4\,\text{s}$, which is
commonly interpreted as magnetic spin-down radiation from an
(in)definitely stable NS formed in a BNS merger (referred to
as the ``magnetar model''; \citealt{Metzger2008,Zhang2001}).

While recent observations of high-mass neutron stars
\citep{Demorest2010,Antoniadis2013} indicate a rather stiff equation
of state and make the formation of a long-lived or even stable
NS a likely possibility in the majority of BNS merger events, the existence of
long-lasting, sustained X-ray afterglows challenge, in particular, the
NS-BH progenitor scenario, as a NS cannot be formed in this case. More
severe, however, is the following apparent dichotomy. On the one hand, a
BH-torus system with accretion timescales of less than one second
cannot sustain continuous energy emission on timescales
$\lesssim\!10^4\,\text{s}$. On the other hand, while the magnetar model can
explain the long-lasting X-ray afterglows, it cannot
readily explain how the prompt SGRB should be generated. Despite
attempts to explain the prompt emission in a way that is similar to long
gamma-ray bursts \citep{Bucciantini2012}, numerical simulations of BNS
mergers have not found indications for the formation of a jet when a
remnant NS is formed (e.g., \citealt{Giacomazzo2013}).

Here we propose a new scenario that can solve this dichotomy. This
scenario assumes a BNS merger that produces a long-lived NS, which emits
spin-down radiation and eventually
collapses to a BH-torus system. While the relativistic jet
produced by the BH-torus system can easily drill through, the spin-down emission is
trapped in a photon-pair plasma nebula and an outer shell of matter ejected shortly
after the NS is formed. The resulting delay can effectively reverse
the observation times of the two signals and can explain why
the X-ray emission powered by the NS spin-down is (in part) observed
after the prompt SGRB emission.

The idea that a delayed afterglow emission could solve the problem
has also been proposed in a parallel work by \citet{Rezzolla2014}, although with
a different phenomenology.\footnote{This idea was originally discussed
  during the preparation of \citet{Siegel2014a} among the three authors,
  but since then it has been independently developed in divergent ways.}    
In this Letter, we focus on showing that the time delay between
the prompt SGRB and the long-lasting X-ray signal can be large enough to explain the
observed X-ray afterglow durations. This represents a fundamental
validation (not considered by \citealt{Rezzolla2014}) and a necessary
condition to make the scenario worth considering.

%%%%%%%%%%%%%%%%%%%%%%%%%%%%%%%%%%%%%%%%%
\section{Phenomenology}
%%%%%%%%%%%%%%%%%%%%%%%%%%%%%%%%%%%%%%%%%

The merger of two NSs forms a differentially rotating object. 
The time-reversal scenario proposed here requires that the merger
remnant is a supramassive NS, i.e., a NS with mass above the maximum
mass for nonrotating configurations $M_\text{TOV}$, but below the
maximum mass for uniformly rotating configurations $M_\text{max}$,
where $M_\text{max}\approx\!1.15\!-\!1.20~M_\text{TOV}$
\citep{Lasota1996}\footnote{The NS could also be hypermassive, i.e.,
  with mass above $M_\text{max}$, if it migrates below this limit by substantial mass loss before
  differential rotation is removed.}. This implies that the NS 
does not require differential rotation to support it against gravitational collapse, and the latter is prevented by uniform rotation as long as the spin rate is high enough; therefore, the time of collapse to a BH is typically of the order of the spin-down timescale.
Given that the mass distribution of BNSs peaks around $1.3\!-\!1.4\,M_\odot$ \citep{Belczynski2008} and that recent observations find NS masses as large as $\simeq\!2\,M_\odot$ \citep{Demorest2010,Antoniadis2013}, supramassive NS merger remnants are a very likely possibility.

In addition to mass ejection associated with the merger
process, the newly born NS can launch baryon-loaded winds
(see Figure~\ref{fig:phenomenology}, (I)), either 
produced by large magnetic pressure gradients at the stellar surface
due to magnetic field amplification in the differentially rotating
stellar interior \citep{Siegel2014a, Siegel2013, Kiuchi2014}, or
induced by neutrino emission (e.g., \citealt{Dessart2009}). Both mechanisms can typically produce mass-loss rates of $\dot{M}\sim 10^{-3}\,M_\odot\,\text{s}^{-1}$ \citep{Dessart2009,Siegel2014a}.
While baryon pollution resulting from dynamical merger ejecta (e.g., \citealt{Hotokezaka2013, Rosswog2013}) is mostly restricted to the orbital plane, the magnetically and neutrino-induced winds are highly isotropic, with bulk speeds $\sim\!0.1~c$ \citep{Siegel2014a}.  
Magnetically induced winds are associated with differential rotation, which can only last for at most $t_\text{dr}\sim 0.1\!-\!10\,\text{s}$, assuming typical initial magnetic field strengths $B\sim 10^{13}\!-\!10^{15}\,\text{G}$ \citep{Shapiro00,Siegel2014a}. Note that during this timescale the initial magnetic fields can be amplified by orders of magnitude. For neutrino-induced winds, the typical cooling timescales are $\lesssim\!1\,\text{s}$ \citep{Dessart2009}; therefore, they can also last no longer than $\sim\!t_\text{dr}$.

At $t\sim t_\text{dr}$, the merger remnant has settled down to a
uniformly rotating, strongly magnetized NS and further mass ejection
is suppressed. As the baryon density in the surrounding of the NS
drops, the NS starts to emit electromagnetic (EM) spin-down radiation at the expense of rotational energy, with luminosity $L_\text{sd}=L_\text{sd}^\text{in}\left(1+t/t_\text{sd}\right)^{-2}$, where
\begin{equation}
L_\text{sd}^\text{in}\simeq1.5\times10^{49}B_{\mathrm{p},15}^2R_6^3P_{\mathrm{in},-3}^{-4}~\mathrm{erg~s^{-1}}.
\label{eq:sd}
\end{equation}
Here, $B_\mathrm{p}$ and $P_\mathrm{in}$ are the surface magnetic field
strength and initial spin period at $t\sim t_\text{dr}$, respectively,
$R$ is the NS radius, and 
\begin{equation}
t_\text{sd}\simeq 2.7\times10^3B_{\mathrm{p},15}^{-2}R_6^{-3}
P_{\mathrm{in},-3}^{2}\,\text{s} \label{eq:t_sd}
\end{equation}
is the spin-down timescale.

The spin-down emission inflates a photon-pair plasma nebula
(henceforth ``nebula"; \citealt{Lightman1987}) behind the expanding ejecta
(see Figure~\ref{fig:phenomenology}, (II)). The high photon pressure
associated with this photon-pair plasma drives a strong shock through
the ejecta, which sweeps up the material into a thin shell while
rapidly propagating toward the outer ejecta radius. 
At shock breakout, a transient signal observed as an early precursor 
to the SGRB \citep{Troja2010} could be produced. Nebula energy is
deposited in the optically thick ejecta and converted into thermal and
kinetic energy, causing a rapid acceleration of the ejecta to
relativistic speeds $v_\text{ej}\lesssim0.8\!-\!0.9~c$ \citep{Metzger2014}.

The injection of spin-down energy into the nebula ceases at
$t=t_\text{coll}$, when rotation can no longer prevent the NS from collapsing
to a BH. Within millisecond timescales, a BH-torus system forms
and generates the necessary conditions to launch a relativistic jet of
duration $\lesssim0.01\!-\!1\,\text{s}$, which corresponds to the
typical accretion time of the torus (e.g.,
\citealt{Narayan92,Shibata06b,Rezzolla:2011}). This jet drills
through the ejecta shell and eventually breaks out, producing the
prompt SGRB emission (see Figure~\ref{fig:phenomenology}, (III)).

As the nebula and the ejecta shell are optically thick for the times of interest, 
the spin-down energy radiated away by the NS up to $t_\text{coll}$ 
emerges from the outer radius of the ejecta shell with
a substantial delay, producing the observed long-lasting X-ray
afterglow radiation. Compared to this delay, the timescale for the jet
to drill through the ejecta and to break out is orders of magnitudes
smaller (see Section~\ref{sec:timing_argument}). Hence, the
spin-down emission, although produced before the jet is formed, can be
observed for a long time after the prompt gamma-ray emission.

An important point to stress is that in the time-reversal scenario the
observed X-ray afterglow durations correspond to such a delay and
not to the spin-down timescale of the NS.

%%%%%%%%%%%%%%%%%%%%%%%%%%%%%%%%%%%%%%%%%
\section{Timing argument}
%%%%%%%%%%%%%%%%%%%%%%%%%%%%%%%%%%%%%%%%%
\label{sec:timing_argument}

The crucial validation to assess whether the time-reversal scenario
outlined above can be compatible with the combined observations of
SGRB prompt emission and long-lasting X-ray afterglows comes from the
analysis of the photon diffusion timescales associated with the nebula and the ejecta.
The scenario cannot hold unless the delay of the signal produced by
the NS spin-down just before the collapse can account for
the observed duration of the X-ray afterglow.

The argument proceeds as follows. From observations (e.g.,
\citealt{Rowlinson2013}) we can 
assume that the NS exists for a time $t_\text{coll}\gtrsim
t_\text{sd}$. Furthermore, we typically have
$t_\text{sd}\gg t_\text{dr}+\Delta t_\text{shock}\equiv t_\text{shock,out}$,
where $\Delta t_\text{shock}$ is the time
needed by the shock to propagate outward through the
ejecta. Consequently, at $t_\text{coll}$ the ejecta has already been
compressed into a thin shell of thickness $\Delta_\text{ej}$, which
moves outward at relativistic speeds.

Most of the emission from the NS cannot reach the outer
radius $R_\text{ej}$ of the ejecta shell later than
$t_\text{NS,out}=t_\text{coll}+\Delta
t_\text{NS,out}$, where $\Delta t_\text{NS,out}$ is
the total time of photon diffusion across the nebula and the ejecta shell.
At $t_\text{NS,out}$, any photon that was present in the system at $t_\text{coll}$ has had
enough time to escape; therefore, the emission is
suppressed for $t\gtrsim t_\text{NS,out}$. The corresponding
delay with respect to the light travel time is given by
$t_\text{NS}^\text{delay}=\Delta t_\text{NS,out}-
R_\text{ej}(t_\text{NS,out})/c$.

The emission from the jet is also delayed due to the time to launch
the jet after the collapse ($\sim\!\text{ms}$) and the fact that the
jet has to propagate outward through the nebula and the ejecta shell
with an effective speed smaller than $c$. While the propagation speed
through the baryon-poor nebula is very close to $c$, the drill time through
the ejecta shell dominates the total delay $t_\text{jet}^\text{delay}$.

Assuming a non-relativistic jet head speed, the drill time across the ejecta can be estimated by \citep{Bromberg2011}
\begin{equation}
  t_\text{drill}\simeq2.5\times10^{-4}\Delta_{\text{ej},9}^{5/3}~\rho_{\text{ej},-7}^{1/3}~(\theta_\text{jet}/30^\circ)^{1/3}~L_{\text{jet},47}^{-1/3}\;\text{s},
\label{eq:tdrill}
\end{equation}
where $\rho_\text{ej}$ denotes the density of the ejecta shell,
$\theta_\text{jet}$ the jet opening angle, and $L_\text{jet}$ the jet
luminosity. For typical parameter values, however,
Equation~\eqref{eq:tdrill} yields a drill time smaller than the corresponding light travel time, which indicates that
the jet head speed has to be relativistic. This is mainly due to the very low
densities at $t\gtrsim t_\text{coll}$ (see Section~\ref{sec:difftimescales}). Consequently,
$t_\text{jet}^\text{delay}$ is negligible with respect to the other
timescales of interest.

In conclusion, a necessary condition to explain a certain duration of the X-ray
afterglow, $\Delta t_\text{afterglow}$, is given by
\begin{equation}
  t_\text{NS}^\text{delay}\simeq t_\text{NS}^\text{delay}-t_\text{jet}^\text{delay}\gtrsim\Delta
  t_\text{afterglow}.\label{eq:timing_condition}
\end{equation}

%%%%%%%%%%%%%%%%%%%%%%%%%%%%%%%%%%%%%%%%%
 \section{Computation of diffusion timescales}
%%%%%%%%%%%%%%%%%%%%%%%%%%%%%%%%%%%%%%%%%
\label{sec:difftimescales}

We compute the diffusion timescales and, hence,
$t_\text{NS}^\text{delay}$ in terms of the following
parameters: the timescale for removal of differential rotation
$t_\text{dr}$, the mass ejection rate $\dot{M}$ (as long as
differential rotation is sustained, i.e., for $t<t_\text{dr}$), the
magnetic field strength at the pole, $B_\mathrm{p}$, and the
initial rotation period $P_\text{in}$ of the NS once it has settled
down to uniform rotation. 
Moreover, we need to specify the time $t_\text{shock,out}$ at
which the shock reaches the outer ejecta radius $R_\text{ej}$, the
expansion speed of the ejecta before and after $t_\text{shock,out}$,
$v_\text{ej}^0$ and $v_\text{ej}$, respectively, 
the pair yield in the nebula, $Y$,
and the ejecta opacity $\kappa$. For the last two parameters, we assume
canonical values $Y\sim 0.1$ and $\kappa\sim 0.2~\mathrm{cm^2~g^{-1}}$
\citep{Metzger2014}. 

At $t=t_\text{dr}$, the NS is still surrounded by a baryon-loaded and
nearly isotropic wind that extends up to
$R_\text{ej}(t_\text{dr})=v_\text{ej}^0t_\text{dr}$ and has an average
density of $\rho_\text{ej}(t_\text{dr})=3t_\text{dr}\dot{M}/[4\pi
R_\text{ej}^3(t_\text{dr})]$. Once differential rotation is removed,
mass ejection is suppressed and spin-down emission drives a strong
shock through the ejecta that rapidly ($t_\text{shock,out}\sim t_\text{dr}$)
sweeps up the ejecta mass ($M_\text{ej}\simeq t_\text{dr}\dot{M}$) into
a thin shell of shocked fluid with thickness $\Delta_\text{ej}$. From
the Rankine-Hugoniot conditions for a strong shock and assuming an
ideal fluid equation of state for the ejecta with $\Gamma=4/3$,
shock compression produces a jump in density of a factor of 7, i.e., at $t=t_\text{shock,out}$:
\begin{equation}
\rho_\text{ej}=\frac{M_\text{ej}}{4\pi R_\text{ej}^2\Delta_\text{ej}}=\frac{21}{4\pi}\frac{\dot{M}t_\text{dr}}{R_\text{so}^3},
\label{eq:rhoej}
\end{equation}
where $R_\text{so}\equiv
R_\text{ej}(t_\text{shock,out})$. This allows us to compute the corresponding
shell thickness $\Delta_\text{ej}=R_\text{so}/21$, which we assume to
be constant for $t>t_\text{shock,out}$. 
After $t_\text{shock,out}$, the ejecta shell is rapidly
accelerated to its asymptotic speed $v_\text{ej}$. The density of the
shell for $t\gg t_\text{shock,out}$ is thus given by
\begin{equation}
  \rho_\text{ej}(t)=\frac{21}{4\pi}\frac{\dot{M}t_\text{dr}}{R_\text{so}[R_\text{so}+v_\text{ej}(t-t_\text{shock,out})]^2}.
\end{equation}

If the ejecta shell at time $t \gg t_\text{shock,out}$ was at rest,
the associated ``static'' photon
diffusion timescale would be
\begin{equation}
  t_\text{diff}^\text{ej,
    stat}(t)=\frac{\Delta_\text{ej}}{c}(1+\kappa\rho_\text{ej}(t)\Delta_\text{ej}),
\label{eq:tdiff_ej}
\end{equation}
where we have added the light travel time to account for a possible
transition to the optically thin regime. Similarly, if the nebula was
not expanding, the associated diffusion timescale resulting from the
electron--positron pairs would be given by \citep{Lightman1987} 
\begin{equation}
  t_\text{diff}^\text{n,stat}(t)=\frac{R_\text{n}(t)}{c}\left(1+\sqrt{\frac{4Y\sigma_\text{T}L_\text{sd}(t)}{\pi
        R_\text{n}(t)m_\text{e}c^3}}\right),\label{eq:tdiff_n}
\end{equation}
where $\sigma_\text{T}$ denotes the Thomson cross section,
$m_\text{e}$ the electron mass, $L_\text{sd}$ the spin down
luminosity, and $R_\text{n}(t)\equiv R_\text{ej}(t)-\Delta_\text{ej}$
the radius of the nebula. 
Note that for $t>t_\text{sd}$, both $t_\text{diff}^\text{ej,stat}$ and
$t_\text{diff}^\text{n,stat}$ are monotonically decreasing functions of time.

The actual photon diffusion timescales differ from
Equations~\eqref{eq:tdiff_ej} and \eqref{eq:tdiff_n}, as the properties of
the system (e.g., $R_\text{ej}$ and $\rho_\text{ej}$) can significantly change while a photon is
propagating outward. Nevertheless, Equations~\eqref{eq:tdiff_ej} and
\eqref{eq:tdiff_n} can be employed to derive lower and upper
bounds. For a photon emitted by the NS just before $t_\text{coll}$, an upper
bound on the total diffusion time through the nebula and the ejecta is
given by the sum of the two static diffusion times at $t_\text{coll}$
(thanks to the monotonicity property mentioned above):  $t_\text{diff}
\lesssim
t_\text{diff}^\text{ej,stat}(t_\text{coll})+t_\text{diff}^\text{n,stat}(t_\text{coll})$. Hence,  
\begin{equation}
  t_\text{NS}^\text{delay}\lesssim t_\text{diff}^\text{ej,stat}(t_\text{coll})+t_\text{diff}^\text{n,stat}(t_\text{coll})-R_\text{ej}(t_\text{coll})/c. \label{eq:t_pul_delay_high}
\end{equation}
A lower bound on $t_\text{NS}^\text{delay}$ can be placed by 
\begin{equation}
  t_\text{NS}^\text{delay}\gtrsim t_\text{diff}^\text{ej,stat}(t_\text{coll}^*)+t_\text{diff}^\text{n,stat}(t_\text{coll}^*)-R_\text{ej}(t_\text{coll}^*)/c,\label{eq:t_pul_delay_low}
\end{equation}
where $t_\text{coll}^*$ is given by $t_{\text{coll}}^*=t_\text{coll}+t_\text{diff}^\text{ej,stat}(t_\text{coll}^*)+t_\text{diff}^\text{n,stat}(t_\text{coll}^*)$.
As a conservative estimate of $t_\text{NS}^\text{delay}$ to be used
in checking the validity of
Equation~\eqref{eq:timing_condition}, we henceforth use the lower
limit given by Equation~\eqref{eq:t_pul_delay_low}.
\begin{figure}[t]
\centering 
\includegraphics[angle=0,width=0.48\textwidth]{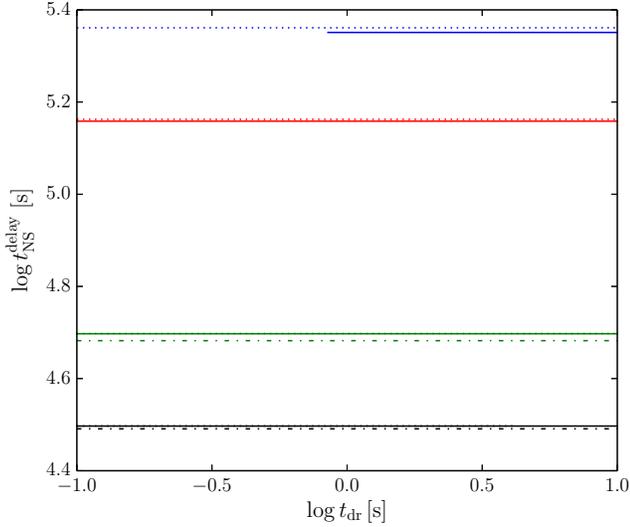}
\caption{Estimated delay for a photon emitted just
  before collapse as a function of $t_\text{dr}$, for
  $B_\text{p}=5\times10^{14},10^{15},5\times 10^{15},10^{16}$\,G
  (from top to bottom) and $P_\text{in}=0.5,1,5$\,ms
  (dotted, solid, dash-dotted), assuming $v_\text{ej}^0=0.1c$, $v_\text{ej} = 0.5c$,
  $\dot{M}=10^{-3}M_\odot\,\text{s}^{-1}$, $t_\text{coll}=t_\text{sd}$. Shown are only cases in which the ejecta shell is still optically thick.}
\label{fig:delay_tdr} 
\end{figure}

\begin{figure}[t]
\centering 
\includegraphics[angle=0,width=0.48\textwidth]{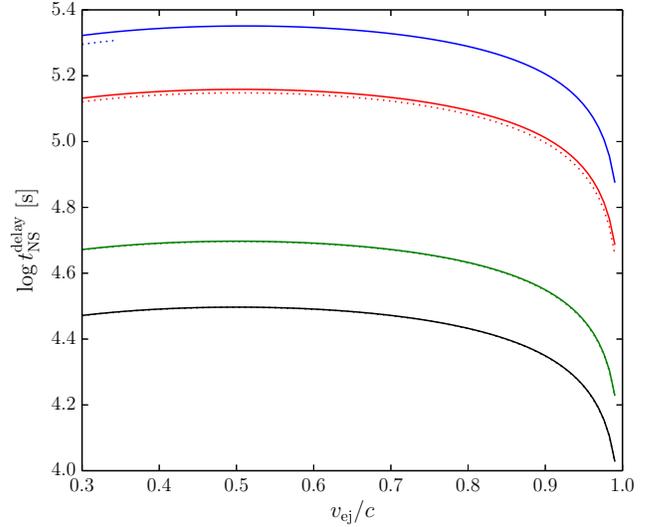}
\caption{Estimated delay for a photon emitted just
  before collapse as a function of the shell expansion speed $v_\text{ej}$, for
 $B_\text{p}=5\times10^{14},10^{15},5\times 10^{15},10^{16}$\,G
  (from top to bottom) and $t_\text{coll}=t_\text{sd}$ (solid),
  $t_\text{coll}=5t_\text{sd}$ (dotted), assuming
  $v_\text{ej}^0=0.1c$, $t_\text{dr}=10\,\text{s}$,
  $\dot{M}=10^{-3}M_\odot\,\text{s}^{-1}$,
  $P_\text{in}=1\,\text{ms}$. Shown are only cases in which the ejecta
  shell is still optically thick.}
\label{fig:delay_vej} 
\end{figure}

Figures \ref{fig:delay_tdr} and \ref{fig:delay_vej} summarize
results for $t_\text{NS}^\text{delay}$ over the
entire parameter space. Reliable estimates can only be provided as
long as the ejecta shell is still optically thick at
$t_\text{coll}^*$ and thus confines the nebula. 
Figure \ref{fig:delay_tdr} shows $t_\text{NS}^\text{delay}$ for a
range of $t_\text{dr}$, $B_\text{p}$, and $P_\text{in}$. The optical
depths decrease for decreasing values of $B_\text{p}$ and increasing values of
$P_\text{in}$, as the spin-down timescale and thus $t_\text{coll}$
increase (see Equation~\eqref{eq:t_sd}). In particular, the ejecta has more
time to expand and thus becomes optically thinner. 
The results are not shown for the cases in which the ejecta shell is optically thin.
As long as the ejecta are optically thick, $t_\text{NS}^\text{delay}$ is
insensitive to the value of $t_\text{dr}$ in the range $\sim\!0.1\!-\!10\,\text{s}$. We have further verified that the same holds for expected
mass-loss rates $\dot{M}\sim\!10^{-4}\!-\!10^{-2}M_\odot\,\text{s}^{-1}$,
shock propagation times $\Delta t_\text{shock}\sim0\!-\!100~t_\text{dr}$, and initial wind
speeds $v_\text{ej}^0\sim0.01\!-\!0.1~c$. This is because
the diffusion time of the nebula is much larger than the diffusion
time of the ejecta shell for the times of interest. Hence,
$t_\text{NS}^\text{delay}$ is
largely insensitive to the properties of the shell.

Figure \ref{fig:delay_vej} reports results on
$t_\text{NS}^\text{delay}$ as a function of the two most influential
parameters of our model, the magnetic field strength of the NS, $B_\text{p}$, and
the asymptotic ejecta shell velocity $v_\text{ej}$. For $B_\text{p}\gtrsim
5\times 10^{14}\,\text{G}$, the ejecta shell is still optically thick at
$t_\text{coll}^*$ for $t_\text{coll}\approx t_\text{sd}$ (solid lines). The dotted
curves indicate that for $t_\text{coll}/t_\text{sd}\gtrsim 5$ and low
magnetic field strengths the ejecta matter can become optically
thin. Figure \ref{fig:delay_vej} also shows that the results only become sensitive to
$v_\text{ej}$ for highly relativistic speeds.

From our estimates we conclude that in the
parameter ranges $B_\text{p}\sim 10^{14}\!-\!10^{16}\,\text{G}$,
$P_\text{in}\sim0.5\!-\!5~\text{ms}$, $t_\text{dr}\sim0.1\!-\!10~\text{s}$, $\dot{M}\sim
10^{-4}\!-\!10^{-2}M_\odot\,\text{s}^{-1}$, $v_\text{ej}^0\sim0.01\!-\!0.1~c$, the delay times
$t_\text{NS}^\text{delay}$ are larger than $3\times 10^{4}\,\text{s}$
and larger than $\sim\!10^{5}\,\text{s}$ for
$B_\text{p}\lesssim 10^{15}\,\text{G}$. Therefore, according to our
criterion given in Equation~\eqref{eq:timing_condition}, the observed
durations of X-ray afterglows of $\Delta t_\text{afterglow}\sim 10^2\!-\!10^5\,\text{s}$ (e.g., \citealt{Rowlinson2013,Gompertz2014}) are compatible with
the proposed time-reversal scenario.

%%%%%%%%%%%%%%%%%%%%%%%%%%%%%%%%%%%%%%%%%
\section{Discussion}
%%%%%%%%%%%%%%%%%%%%%%%%%%%%%%%%%%%%%%%%%

In our scenario, the long-lasting
($\sim\!10^2\!-\!10^5~\mathrm{s}$) X-ray afterglow emission accompanying
a large fraction of SGRBs is
produced by a long-lived supramassive NS, which eventually
collapses to a BH on the spin-down timescale and thus generates the
necessary conditions for a relativistic jet to be launched. 
While the jet can easily drill through the surrounding photon-pair
plasma and baryon-loaded ejecta, spin-down radiation emitted by the NS
before the collapse diffuses outward on a much longer timescale,
accumulating a significant delay before finally escaping. This delay
explains how the X-ray emission powered by spin-down radiation before
the collapse is observed in part before and in part after the prompt emission. 
The optically thick nebula and ejecta are therefore responsible for a
``time reversal'' of the observed signals. 

In our simple analysis, we have focused on the estimation of the maximum
delay that can affect the spin-down radiation, depending on the
most relevant properties of the system. If this delay is shorter
than the duration of an observed X-ray afterglow, the latter cannot be
explained within the time-reversal scenario. Therefore, this test is
crucial to make the scenario worth considering.

By exploring a wide range of physical parameters, we find that in most
cases, afterglow durations of up to $10^4\!-\!10^5~\mathrm{s}$ are compatible
with the estimated delays. Moreover, we find that the maximum delay is
always determined by the high optical depth of the nebula, which
dominates over the optical depth of the ejecta. 

One important consequence of this new scenario is that for all
SGRB events accompanied by a long-lasting X-ray afterglow, the
progenitor would be necessarily a BNS and not a BH-NS
binary. 
Furthermore, the peak amplitude of GW emission associated with the
time of the merger would reach the observer long before the EM prompt
SGRB signal. The two
signals would be separated by the lifetime of the supramassive NS,
which can easily exceed $\sim\!10^3~\mathrm{s}$, with profound
consequences for coincident GW and EM observations. 
This separation would provide an accurate measurement of 
the NS lifetime (to better than 1\%).

A supporting piece of evidence for our scenario comes from the observation of
steep decay phases in some of the X-ray afterglow light curves, which
are interpreted as a collapse to a BH (e.g., \citealt{Rowlinson2013}). These
features are in favor of the long-lived supramassive NS assumed here.
A strong indication for the time-reversal scenario would come from the observation
of X-ray ``afterglow'' emission prior to the SGRB itself, i.e., of a
plateau-like X-ray light curve with the prompt emission in between. 
This would be indicative of a change in arrival times with respect to
the emission times of the two signals.
Moreover, the observation of ``orphan" SGRB-like events, with a
plateau-like X-ray emission, but missing prompt emission, can confirm
the isotropy of the afterglow radiation we expect. In this case, the collimated
prompt emission would be beamed away from us.

A potential difficulty is posed by the observation of several SGRB
events in which the X-ray afterglow light curve shows a late-time decay
compatible with $L_\text{X}\propto t^{-2}$. 
This is the behavior expected for spin-down radiation at $t>t_\text{sd}$
(see above Equation~(\ref{eq:sd})). In our scenario, we need to explain how this
particular decay is not altered by the optically thick environment 
surrounding the NS. We find that
for the cases considered here, the delay due to the diffusion time of
photons through the ejecta shell at times $t\gtrsim t_\text{sd}$ is typically
orders of magnitude smaller than the spin-down timescale and, thus, such
diffusion should have no effect on the time behavior of the signal
luminosity.
The diffusion timescale associated with the nebula, however, is
not smaller than the spin-down timescale and whether the trapped
radiation would still emerge with the same time
dependence as the spin-down injection luminosity remains in doubt.  

\bigskip
We thank B. Schutz and W. Kastaun for valuable discussions. R.C.
acknowledges support from MIUR FIR grant No. RBFR13QJYF. 
 
%%%%%%%%%%%%%%%%%%%%%%%%%%%%%%%%%%%%%%%%%

%\bibliographystyle{apj}
%\bibliography{aeireferences}

%%%%%%%%%%%%%%%%%%%%%%%%%%%%%%%%%%%%%%%%%
\end{document}